\documentclass[runningheads,a4paper]{llncs}
\usepackage[misc]{ifsym}
\usepackage{amssymb}
\usepackage[colorlinks,urlcolor=blue,citecolor=blue]{hyperref}
\usepackage{graphicx}
\usepackage{booktabs}
\usepackage{url}
\newcommand{\keywords}[1]{\par\addvspace\baselineskip
\noindent\keywordname\enspace\ignorespaces#1}
\usepackage{bbding}

\begin{document}

\mainmatter  

\title{Provably Secure Integration Cryptosystem on Non-Commutative Group}

\titlerunning{Provably Secure Integration Cryptosystem on Non-Commutative Group}

%
%
            \author{Xiaoming Chen$^1$, Weiqing You$^2$%
            }
            \authorrunning{X. Chen et al}

            \institute{$^{1,2}$ Beijing Electronic Science \& Technology Institute
            Beijing 100070, China\\
            $^1$ University of Science and Technology of China, Hefei 230026, China \\
            chenxmphd@yeah.net, scipaperyou@sina.com}

%
%

\toctitle{Lecture Notes in Computer Science}
\tocauthor{Authors' Instructions}
\maketitle

\begin{abstract}
Braid group is a very important non-commutative group. It is also an important tool of quantum field theory, and has good topological properties. This paper focuses on the provable security research of cryptosystem over braid group, which consists of two aspects: One, we prove that the Ko's cryptosystem based on braid group is secure against chosen-plaintext-attack(\emph{CPA}) which proposed in CRYPTO 2000, while it dose not resist active attack.  The other is to propose a new public key cryptosystem over braid group which is secure against adaptive chosen-ciphertext-attack(\emph{CCA2}). Our proofs are based on random oracle models, under the computational conjugacy search assumption(\emph{the CCS assumption}). This kind of results have never been seen before.
\end{abstract}

\section{Introduction}
\subsection{Background and related work}

In 1994, Shor\cite{jour2} proposed a quantum fourier transforma algorithm, which can construct an integer factorization polynomial (quantum) algorithm, which poses a substantial threat to the security of RSA. In 2003, Proos et al. \cite{jour3} extended the Shor algorithm to the elliptic curve, and obtained the polynomial (quantum) algorithm for solving the discrete logarithm problem over the elliptic curve, which poses a substantial threat to the security of ECC. However, these famous quantum algorithms mainly focus on the exchange structure. For some Cryptosystems based on noncommutative structures\cite{jour1}\cite{jour4}\cite{jour5}\cite{jour6},that attacks are ineffective. Therefore, the design of cryptographic systems over certain non-commutative groups is one of the most important way to find algorithms which can resist quantum attacks. It is the key research object in the field of post quantum cryptography.

The braid group is a very important infinite non-commutative generation group. Because of its many difficult problems and many commutative subgroups, that make it can be used as the carrier of the design of cryptographic systems. The braid group has good algebraic properties, making it a good platform for designing quantum attack algorithms. In 2000, Ko et al proposed a public key cryptosystem based on braid group, after that, there are many papers about the design of the braid cryptosystem in \cite{jour8}\cite{jour9}\cite{jour10}, followed by some questions about the hypothesis of the braid base problem \cite{jour9}\cite{jour11}\cite{jour12}\cite{jour13}\cite{jour14} were proposed. However, as far as the existing technology and theory are concerned, the conjugate problem on the braid group is still difficult\cite{jour16}\cite{jour17}, that is, there is no polynomial algorithm that can solve the conjugate problem on the braid group in polynomial time, and even in quantum computation, there is no effective algorithm for the conjugate problem at present.

After many years of research and development, people have a deeper understanding of braid cryptology, especially the starting point of the braid group, which greatly promotes the research of cryptographic systems on noncommutative group\cite{jour33}\cite{book1}\cite{book2}. On the other hand, there are some fast computation algorithms were proposed\cite{jour25}\cite{jour28}\cite{jour34}\cite{jour35}, and the implementation of this algorithm has been solved by the center of steven research on algebraic\cite{url}.
Recently, there are some digital signature algorithms were proposed, such as WalnutDSA\cite{jour50}\cite{jour51}\cite{jour52}, and others schemes was proposed\cite{jour18}\cite{jour19}. These scheme are very attractive. The performance of computing and storage is approaching the need of application.

But so far, the research on the proof security of braid cryptosystems is very rare or even empty, which greatly hinders the delovepment and application of braid cryptosystem. The security of IND-CPA, IND-CCA and IND-CCA2 is enhanced in turn \cite{jour20} \cite{jour21} \cite{jour22}, and the structure of cipher algorithm is becoming more and more complex, and the consumption of computation is also increasing. The early construction of CCA or CCA2 security is realized by the zero knowledge proof method, so the cryptographic algorithm constructed is very practical. In 1993, Bellare and Rogaway\cite{jour23} proposed a method to prove IND-CCA2 under the random oracle model. The model is concise and is widely recognized and loved by the researchers. Although the security conclusion of the cryptographic algorithm in this model does not fully represent the actual security\cite{jour23}, it is still the most effective index of security. The ROM model and method are still the main technology of the public key cryptographic security argument. The public key cryptography algorithm based on braid group also uses ROM model to prove security.

\subsection{Our result}

There are more detailed studies on the definition, basic concepts and computational methods of the braid group\cite{jour24}, this article will not be described here. But the main section is focused on the proof of security of the braid group cryptography algorithm. Our main work is as follows:

$1.$ We have finished the research on the indistinguishability of the braid cryptosystem proposed by Ko\cite{jour1}, proved that it is IND-CPA through the random oracle model under the computational conjugacy search assumption(the CCS assumption), and we emphasize that it does not have the ability to resist active attack.

$2.$ According to the original EIGamal scheme design idea, we propose a cryptographic algorithm with IND-CPA security under the standard model and the decisional conjugacy search assumpiton(the DCS assumption).

$3.$ Adopting the design idea of hybrid encryption system, we propose a new public key cryptosystem in braid against adaptive chosen ciphertext attack. Subsequently, its IND-CCA2 security is proved under the random oracle model.

Before this paper, there is no any research on the provable secure encryption algorithm on braid group, Our algorithm and proof fill this gap. Like all the provable security analysis procedures, the proof part of this article has taken a lot of space, but its logical process is not very complicated.

\section{Preliminaries}

\subsection{Braid Group}

Compared with the general group, the structure of the braid group is more special and complicated. Although the introduction of braid group theory has been very detailed, we still need to spend some words to introduce the basic theories related to it. If the readers need more about braid theory, please refer to literature\cite{jour24}

\subsubsection{Definition 1}

Define $B_n$ as a braid group generated by $\sigma_1, \sigma_2, \cdots, \sigma_{n-1}$, and following the relations:

\begin{displaymath}
    \left\{ \begin{array}{cc}
    \sigma_i\sigma_j\sigma_i=\sigma_j\sigma_i\sigma_j & \textrm{    \indent if $|i-j|=1$}\\
    \sigma_i\sigma_j=\sigma_j\sigma_i & \textrm{   \indent if $|i-j| \geq 2$} \\
    \end{array} \right.
\end{displaymath}

The string formed by generators in braid group $B_n$ is called a braid (or a word), and the number of generators in the string is the length of the braid (or word). It can be clearly seen that the braid group is a class of non commutative generating groups, but there are a large number of commutative elements on it. It is easy to see that there are many commutative subgroups on it. Assume $B_{l+r}=\{\sigma_1, \sigma_2, \cdots, \sigma_{l+r-1} \}$, let $LB_l=\{\sigma_1, \sigma_2, \cdots, \sigma_{r-1} \}$  be a left subgroup, and $RB_r=\{\sigma_{r+1}, \sigma_{r+2}, \sigma_{r+l-1} \}$ be a right subgroup. So
$$ \forall x \in LB_l, \forall y \in RB_r,  xy=yx $$

This is the basis for computing for building an available key exchange protocol and a cryptographic algorithm.

\subsubsection{Definition 2} The fundamental braid is represented by the symbol $\Delta$:
\begin{displaymath}
    \left\{ \begin{array}{ll}
    \Delta\ =1 \\
    \Delta_n=\Delta_{n-1}\sigma_{n-1}\sigma_{n-2}\cdots\sigma_1
    \end{array} \right.
\end{displaymath}

\subsubsection{Theorem 1}\cite{jour25} Every word $w$ in braid can be represented as a canonical form: $W=\Delta^kA, k\in Z, A \in B_n^+$, or the canonical form for short. Of course, there are many standard forms of it. Please refer to the literature\cite{jour26}\cite{jour27}\cite{jour28}\cite{jour29}. For the sake of convenience, this paper adopts the left canonical form.

The literature\cite{jour1} enumerated 7 hard problems in the braid group, we show that problems related to this paper as follow: \\

\noindent 1. Conjugacy Decision Problem \\
\indent \textbf{Instance:} $(x,y) \in B_n \times B_n.$ \\
\indent \textbf{Objuctive:} Determine whether $x$ and $y$ are conjugate or not. \\

\noindent 2. Conjugacy Search Problem \\
\indent \textbf{Instance:} $(x,y) \in B_n \times B_n$ such that $x$ and $y$ are conjugate. \\
\indent \textbf{Objuctive:} Find $a \in B_n $ such that $y=axa^{-1}$. \\

\noindent 3. Generalized Conjugacy Search Problem \\
\indent \textbf{Instance:} $(x,y) \in B_n \times B_n$ such that $y=axa^{-1}$ for some $b\in B_m$, $m \leq n $.  \\
\indent \textbf{Objuctive:} Find $a \in B_m $ such that $y=axa^{-1}$. \\

These hard problems are very useful for the analysis of public key cryptosystems,thus, we will use them to construct the security assumption.

\subsection{Security Model}
The security model is portrayed by Indistinguishability-Game (IND-GAME), mainly divided into three levels: Indistinguishability-Chosen Plaintext Attack (IND-CPA) \cite{jour20}, Indistinguishability - (Non Adaptive) Chosen Ciphertext Attack (IND-CCA) \cite{jour21}, Indistinguishability - (Adaptive) Chosen Ciphertext Attack (IND-CCA2) \cite{jour22}.

\subsubsection{Definition 3 Indistinguishability-Chosen Plaintext Attack (IND-CPA)}
The IND game of public key encryption scheme under chosen plaintext attack (IND-CPA) is as follows\cite{jour20}:\\

Step1. Initialization﹝ The Challenger $B$ generates the password system, and the Adversary $A$ obtains the system public key $pk$.

Step2. The Adversary $A$ generates plaintext messages and obtains encrypted ciphertext (polynomial bounded).

Step3. Challenge. The Adversary $A$ outputs two messages of the same length, $M_0$ and $M_1$. The Challenger $B$ chooses $\beta\leftarrow_R\{0,1\}$, cipher $M_{\beta}$, and sends ciphertext $C^\ast$ (Target ciphertext) to $A$.

Step4. Guess. $A$ outputs $\beta'$, if $\beta'=\beta$, return 1, $A$ attack successfully.

The advantage of the adversary $A$ can be defined as a function of the parameter $K$:
$$ Adv_A^{CPA}(K)=\left| Pr[\beta'=\beta]-\frac{1}{2} \right| $$
For a polynomial time adversary $A$, there is a negligible function $\varepsilon(K)$ that makes $Adv_A^{CPA}(K)\leq \varepsilon(K)$ set up, it is called IND-CPA security.

\subsubsection{Definition 4 Indistinguishability - (Non Adaptive) Chosen Ciphertext Attack (IND-CCA)}\cite{jour21}
The IND game of public key encryption scheme under chosen ciphertext attack (IND-CCA) is as follows\cite{jour21} \\

Step1. Initialization﹝ The Challenger $B$ generates the password system, and the Adversary $A$ obtains the system public key $pk$.

Step2. Training. $A$ sends the ciphertext $C$ to the $B$, and $B$ sends the decrypted plaintext to $A$.(Polynomial bounded)

Step3. Challenge. The Adversary $A$ outputs two messages of the same length, $M_0$ and $M_1$. The Challenger $B$ chooses $\beta\leftarrow_R\{0,1\}$, cipher $M_{\beta}$, and sends ciphertext $C^\ast$ (Target ciphertext) to $A$.

Step4. Guess. $A$ outputs $\beta'$, if $\beta'=\beta$, return 1, $A$ attack successfully.

The advantage of the adversary $A$ can be defined as a function of the parameter $K$:
$$ Adv_A^{CCA}(K)=\left| Pr[\beta'=\beta]-\frac{1}{2} \right| $$
For a polynomial time adversary $A$, there is a negligible function $\varepsilon(K)$ that makes $Adv_A^{CCA}(K)\leq \varepsilon(K)$ set up, it is called IND-CCA security.

The above attack is also called 'lunch time attack'. At a 'lunch time', the enemy has a black box that can perform the decryption operation, and the black box can not be used after 'lunch time'.

\subsubsection{Definition 5 Indistinguishability - (Adaptive) Chosen Ciphertext Attack (IND-CCA2)}\cite{jour22}
The IND game of public key encryption scheme under adaptive chosen ciphertext attack (IND-CCA2) is as follows\cite{jour22} \\

Step1. Initialization﹝ The Challenger B generates the password system, and the Adversary $A$ obtains the system public key $pk$.

Step2. Training1. $A$ sends the ciphertext $C$ to the $B$, and $B$ sends the decrypted plaintext to $A$.(Polynomial bounded)

Step3. Challenge. The Adversary $A$ outputs two messages of the same length, $M_0$ and $M_1$. The Challenger $B$ chooses $\beta\leftarrow_R\{0,1\}$, cipher $M_{\beta}$, and send ciphertext $C^\ast$ (Target ciphertext) to $A$.

Step4. Training2. $A$ sends the ciphertext $C(C\neq C^\ast)$ to the $B$, and $B$ sends the decrypted plaintext to $A$.(Polynomial bounded)

Step5. Guess. $A$ outputs $\beta'$, if $\beta'=\beta$, return 1, $A$ attack successfully.

The advantage of the adversary $A$ can be defined as a function of the parameter $K$:
$$ Adv_A^{CCA2}(K)=\left| Pr[\beta'=\beta]-\frac{1}{2} \right| $$
For a polynomial time adversary $A$, there is a negligible function $\varepsilon(K)$ that makes $Adv_A^{CCA2}(K)\leq \varepsilon(K)$ set up, it is called IND-CCA2 security.

\section{Two Schemes Provably Secure Against Chosen Plaintext Attack}
In order to research on the indistinguishable security of public key algorithms over the braid group , we first give the following two assumptions:
\subsubsection{The Compution Conjugacy Search Assumption (The CCS Assumption)} Given $X,Y\in B_n, X=xgx^{-1}, Y=ygy^{-1}$, it is hard to compute
$Z=(xy)g(xy)^{-1}$
\subsubsection{The Decisional Conjugacy Search Assumption (The DCS Assumption)} Assume that $B_{l+r}$ is a braid group, $LB_l$ and $RB_r$  are left subgroup and right subgroup, respectively. Assume $g,z \leftarrow_R B_{l+r}, x \leftarrow_R LB_l, y \leftarrow_R RB_r$, The following two distributions are computationally non - distinguishable:
\begin{eqnarray}
R & = & (g,xgx^{-1},ygy^{-1},zgz^{-1}) \\
D & = & (g,xgx^{-1},ygy^{-1},(xy)g(xy)^{-1})
\end{eqnarray}
We can call the distribution $R$ is Random four tuple while the distribution $D$ is $DCS$ four tuple.
\subsection{A Scheme Provably Secure Against Chosen Plaintext Attack}
Before analyzing Ko. public key cryptosystem\cite{jour1}, we first propose a non hashing braid group public key cryptosystem, which is very similar to the original ELGamal Scheme\cite{jour30}.

\subsubsection{Algorithm 1} Assume $B_{l+r}$ is a braid group, left subgroup $LB_l$ and right subgroup $RB_r$. \\

\noindent \textbf{KeyGeneration} One selects a element $g\leftarrow_R B_{l+r}, x\leftarrow_R LB_l, X=xgx^{-1}$, the public key is $(X, g)$, the private key is $(x,g)$.

\noindent \textbf{Encryption} The cipher gets a message $m\in B_{l+r}$, one selects a element $y\leftarrow_R RB_r$, compute $Y=ygy^{-1}, Z=yXy^{-1}, c=Zm$. The ciphertext is $(Y,c)$.

\noindent \textbf{Decryption} The decipher gets the target ciphertext $(Y,c)$, computes $Z=xYx^{-1}, m=Z^{-1}c$.

\subsubsection{Theorem 2} If the DCS assumption holds, the algorithm 1 is $IND-CPA$.\\

\textbf{Proof}: Assume a PPT adversary $A$ attack algorithm 1, $A$ outputs $M_0,M_1$, the challenger $B$ chooses $\beta\leftarrow_R \{0,1\}$, cipher it and sents the ciphertext to $A$. $A$ runs an randomization algorithm, outputs the guessing value ${\beta}$. If $\beta'=\beta$, A attack successfully, represented by event $succ$. Note that the advantage of $A$ is
$$Adv_A^1=\left| \frac{1}{2} - Pr[Succ] \right|$$
The following constructs an adversary $B$, $B$ uses $A$ to attack the DCS assumption. Assume B output the tuple $T=(g_1,g_2,g_3,g_4)$, the advatage of $B$ is
$$Adv_B^1=\left| \frac{1}{2} - Pr[\beta'=\beta] \right|$$

The structure of the $B$ is shown as follows:

\begin{displaymath}
  \begin{array}{l}
    Experiment_B^1(T): \\
    \indent pk=(g_1, g_2);\\
    \indent (M_0,M_1)\leftarrow A(pk), \left| M_0 \right|=\left| M_1 \right|=l(K);\\
    \indent \beta\leftarrow_R(0,1);\\
    \indent C^\ast=(g_3,g_4M_{\beta});\\
    \indent \beta'\leftarrow A(pk,C^\ast);\\
    \indent If  \quad \beta'=\beta, return \quad  1;\\
    \indent else \quad return \quad 0.
  \end{array}
\end{displaymath}

When return 1, $B$ guesses that the input $T$ is four tuples $DCS$, else $B$ guesses that the input $T$ is random four tuples. Let $R$ represent events '$T$ is the random four tuples', $D$ represent events '$T$ is the DCS four tuples'. Two steps of proof: \\

$1.$ $Pr[Exp_B^1(T)=1|R]=\frac{1}{2}$. \\

When the 'event R' happened, $g_4$ is a random element in $B_{l+r}$, so it is independent of the ciphertext $C^\ast$. Thus, $A$ have no any information of $\beta$, he can't guess $\beta$ with more than 1/2 probability. When $B$ return 1 if and only if $A$ success, so $Pr[Exp_B^1(T)=1|R]=\frac{1}{2}$. \\

$2.$ $Pr[Exp_B^1(T)=1|D]=Pr[Succ]$. \\

When the 'event D' happened, $g_2=xg_1x^{-1}, g_3=yg_1y^{-1}, g_4=yg_2y^{-1}$. So $B$ return 1 if and only if $A$ success.

\begin{eqnarray*}
Pr[Exp_B^1(T)=1] & = & Pr[D]Pr[Exp_B^1(T)=1|D] + Pr[R]Pr[Exp_B^1(T)=1|R] \\
                 & = & \frac{1}{2}Pr[Succ] + \frac{1}{2} \times \frac{1}{2}
\end{eqnarray*}

\begin{eqnarray*}
Pr[Exp_B^1(T)=0] & = & Pr[D]Pr[Exp_B^1(T)=0|D] + Pr[R]Pr[Exp_B^1(T)=0|R] \\
                 & = & \frac{1}{2}(1 - Pr[Succ]) + \frac{1}{2} \times \frac{1}{2}
\end{eqnarray*}

so, $ \left| Pr[Exp_B^1(T)=1] - Pr[Exp_B^1(T)=0] \right| = \left| Pr[Succ] - \frac{1}{2} \right|$ \\

If $A$ attacks $B$ with the non negligible advantage of $\varepsilon(K)$, then $B$ attacks the DCS assumption with the same advantage.

\subsection{The Security of Ko's cryptosystem}
In the provable security theory of public key cryptography, the weaker the security assumption is, the more rigorous the results are. Like the DDH assumption and the CDH assumption\cite{jour31}, The DCS assumption is more stronger than the CCS assumption. So a scheme under the CCS assumption is more security. The following algorithm 2 was proposed by Ko et al. in crypto 2000\cite{jour1}.

\subsubsection{Algorithm 2} Assume $B_{l+r}$ is a braid group, left subgroup $LB_l$ and right subgroup $RB_r$, $H\leftarrow_R \{ H:B_{l+r}\rightarrow \{0,1\}^{l(k)} \}$ is a hash function. \\

\noindent \textbf{KeyGeneration} One selects a element $g\leftarrow_R B_{l+r}, x\leftarrow_R LB_l, X=xgx^{-1}$, the public key is $(X, g)$, the private key is $(x,g)$.

\noindent \textbf{Encryption} The cipher gets a message $m\in B_{l+r}$, one selects a element $y\leftarrow_R RB_r$, computes $Y=ygy^{-1}, Z=yXy^{-1}, c=H(Z) \oplus m$. The ciphertext is $(Y,c)$.

\noindent \textbf{Decryption} The decipher gets the target ciphertext $(Y,c)$, computes $Z=xYx^{-1}, m=H(Z) \oplus c$.

\subsubsection{Theorem 3} If $H$ is a random oracle and the CCS assumption holds, then the algorithm 2 is secure against chosen plaintext attack.

\textbf{Proof}: Assume that $A$ is an $IND-CPA$ adversary who attacks algorithm 2, the advantage of $A$ is $Adv_A^2$, $B$ is a adversary who attacks the CCS assumption, the advantage of $B$ is $Adv_B^2$. If $A$ attacks algorithm 2 with the non negligible advantage of $\varepsilon(K)$, it must exist $B$ whom attacks the CCS assumption with the advantage of $Adv_B^2 \geq 2\varepsilon(K)$ . The $IND-CPA$ game of algorithm 2 is described as follows:

\begin{displaymath}
  \begin{array}{l}
    Exp_A^2(K): \\
    \indent g \leftarrow_R B_{l+r}, x \leftarrow_R LB_l, X=xgx^{-1};\\
    \indent pk=(g, X), sk = x;\\
    \indent (M_0,M_1)\leftarrow A^{H(\cdot)}(pk), \left| M_0 \right|=\left| M_1 \right|=l(K);\\
    \indent \beta\leftarrow_R(0,1), y \leftarrow_R RB_r, Y=ygy^{-1}, Z=yXy^{-1},C^\ast=(Y,H(Z)\oplus M_{\beta}); \\
    \indent \beta'\leftarrow A^{H(\cdot)}(pk,C^\ast);\\
    \indent If  \quad \beta'=\beta, return \quad  1;\\
    \indent else \quad return \quad 0.
  \end{array}
\end{displaymath}

The advantage of $A$ can be define as a function of the security parameter $K$
$$Adv_A^2(K)=\left| Pr[Exp_A^2(K)=1]-\frac{1}{2} \right|$$

$B$ gets $(g, X, \hat{c}_1)$, $\hat{c}_1$ is the first component of the target ciphertext. Using the $A$ attacks algorithm 2 as a subprogram, The following steps are taken to calculate $\hat{y}$, $\hat{c}_1=\hat{y} g \hat{y}^{-1}$. \\

$1.$ One chooses a random string $ \hat{h}\leftarrow_R {0,1}^{l(k)} $ as a guessing value for $H(\hat{Z})$, $B$ don't know the $\hat{Z}$, sents $pk=(g,X)$ to $A$; \\

$2.$ $H$ queries (Bounded polynomial times): $B$ build the list $H^{list}$ (Initial empty), element type is $( \hat{Z}_i, \hat{h}_i )$, $A$ can query $H^{list}$ any time, $B$ respond as follows:

$\bullet$ If $ \hat{Z}$ in $H^{list}$, respond with $\hat{h}$ in $(\hat{Z}, \hat{h})$.

$\bullet$ If $ \hat{c}_1 = \hat{y} X \hat{y}^{-1}$, respond with $\hat{h}$ , record $ \hat{Z} = \hat{y} X \hat{y}^{-1}$, save the $ (\hat{Z},\hat{h}) $ into list.

$\bullet$ Else, choose random string $ \hat{h} \leftarrow_R \{ 0, 1 \}^{l(k)} $, respond with $\hat{h}$, record $ \hat{Z} = \hat{y} X \hat{y}^{-1}$, save the $ (\hat{Z},\hat{h}) $ into list.\\

$3.$ Challenge. $A$ outputs two messages $ M_0, M_1 $, $B$ chooses $\beta\leftarrow_R \{0,1\}$, sets $ \hat{c}_2=\hat{h}\oplus M_{\beta} $, sends $( \hat{c}_1, \hat{c}_2 )$ (as ciphertexts) to $A$;\\

$4.$ After end of above steps, $A$ outputs $\beta'$, $B$ outputs $ \hat{Z} = \hat{y} X \hat{y}^{-1}$ which recorded in step2.

Assume that the $event \ D$: In the simulation, $H(\hat{Z})$ appears in the list $H^{list}$.

\subsubsection{Assertion 1} $B$ is complete in above simulation process.

\textbf{proof}: It is easy to know:

$\bullet$ In the $H$ inquiry of $A$, each value is answered by random string. In the real attacks of $A$, the value of the function is generated by the random oracle, so the function value obtained by the $A$ is uniformly distributed;

$\bullet$ For $A$, $ \hat{h}\oplus M_{\beta} $ is a one-time pad system, From the randomness of $\hat{h}$, it is known that $ \hat{h}\oplus M_{\beta} $ is random for $A$.

So, Both of the view of A and its view in real attacks are not distinguishable in calculation.

\subsubsection{Assertion 2} In the simulation attack above, $Pr[D]\geq 2\varepsilon$ \\

\noindent \textbf{proof} Obviously $Pr[Exp_A^2(K)=1|\neg D]=\frac{1}{2}$, \\

\noindent $ \because \ Adv_A^2 \geq \varepsilon(K)$, in the simulation:
\begin{eqnarray*}
Pr[Exp_A^2(K)=1] & = & Pr[\neg D]Pr[Exp_A^2(K)=1|\neg D] + Pr[D]Pr[Exp_A^2(K)=1|D]  \\
                 & \leq &  Pr[\neg D]Pr[Exp_A^2(K)=1|\neg D] + Pr[D]  \\
                 & = & \frac{1}{2}Pr[\neg D] + Pr[D] \\
                 & = & \frac{1}{2} +\frac{1}{2}Pr[D] \\
Pr[Exp_A^2(K)=1] & \geq & Pr[\neg D]Pr[Exp_A^2(K)=1|\neg D] \\
                 & = & \frac{1}{2}Pr[\neg D] \\
                 & = & \frac{1}{2} - \frac{1}{2}Pr[D]
\end{eqnarray*}
\noindent $ \therefore \ \frac{1}{2} - \frac{1}{2}Pr[D] \leq Pr[Exp_A^2(K)=1] \leq \frac{1}{2} +\frac{1}{2}Pr[D]$ \\

\noindent $ \therefore \  Adv_A^2(K) = \left| Pr[Exp_A^2(K)=1] - \frac{1}{2} \right| \leq \frac{1}{2}Pr[D]$ \\

\noindent $ \therefore \ \frac{1}{2}Pr[D] \geq 2Adv_A^2(K) = 2 \left| Pr[Exp_A^2(K)=1] - \frac{1}{2} \right| \geq 2\varepsilon(K) $ \\

In summary, if $A$ can take advantage of a non negligible advantage $\varepsilon(K)$ to attack algorithm 2, it must exist $B$ whom take $ \ Adv_B^2 \geq 2\varepsilon(K) \ $ to attack the CCS assumption. So the algorithm 2 is secure against $IND-CPA$.

\section{A Public Key Cryptosystem on Braid Provably Secure Against Adaptive Chosen Ciphertext Attack }

Under the current complex network environment, it is entirely possible for the adversary to achieve active attack. Therefore, it is very important for the study of an algorithm with IND-CCA2 security. Next, we will propose a cryptographic algorithm secure against adaptive chosen ciphertext attack on braid groups.

\subsubsection{Algorithm 3} $(E,D)$ is a pair of symmetric key algorithms secure against adaptive chosen ciphertext attack, other conditions are the same as Algorithm 1. \\

\noindent \textbf{KeyGeneration} One selects a element $g\leftarrow_R B_{l+r}, x\leftarrow_R LB_l, X=xgx^{-1}$, the public key is $(X, g)$, the private key is $(x,g)$.

\noindent \textbf{Encryption} The cipher gets a message $m\in B_{l+r}$, one selects a element $y\leftarrow_R RB_r$, computes $Y=ygy^{-1}, Z=yXy^{-1}, k=H(Z), c=E_k(m)$. The ciphertext is $(Y,c)$.

\noindent \textbf{Decryption} The decipher gets the target ciphertext $(Y,c)$, computes $Z=xYx^{-1}, k=H(Z), m=D_k(m) $.

\subsubsection{Theorem 4} If $H$ is a random oracle, and the CCS assumption holds, then the algorithm 3 is secure against chosen ciphertext attack. \\

\noindent \textbf{Proof}: Suppose an IND-CCA2's adversary $A$ breaks the algorithm 3 with a not negligible advantage $\varepsilon(k)$, then there must be an adversary $B$ attacks the CCS assumption with the advantage of $Adv_B^3 \geq 2\varepsilon(k) $. \\

 Assume that $A$ is an IND-CCA2 adversary who attacks algorithm 3, the advantage of $A$ is  $Adv_A^3$, $B$ is a adversary who attacks the CCS assumption, the advantage of $B$ is $Adv_B^3$. If $A$ can take advantage of a non negligible advantage $\varepsilon(K)$ to attack algorithm 3, it must exist $B$ whom takes $Adv_B^3 \geq \varepsilon(K)$ to attack the CCS assumption. The IND-CCA2 game of algorithm 3 is described as follows:

\begin{displaymath}
  \begin{array}{l}
    Exp_A^3(K): \\
    \indent g \leftarrow_R B_{l+r}, x \leftarrow_R LB_l, X=xgx^{-1};\\
    \indent pk=(g, X), sk = x;\\
    \indent (M_0,M_1)\leftarrow A^{H(\cdot),D_{sk}(\cdot)}(pk), \left| M_0 \right|=\left| M_1 \right|=l(K);\\
    \indent \beta\leftarrow_R\{0,1\}, y \leftarrow_R RB_r, Y=ygy^{-1}, Z=yXy^{-1}, k=H(Z), C^\ast=(Y,E_k(M_{\beta}); \\
    \indent \beta'\leftarrow A^{H(\cdot),D_{sk,\neq C^\ast}}(pk,C^\ast);\\
    \indent If  \quad \beta'=\beta, return \quad  1;\\
    \indent else \quad return \quad 0.
  \end{array}
\end{displaymath}

$D_{sk,\neq C^\ast}$ means $A$ can not query $C^\ast$ to the oracle. The advantage of $A$ can be define as a function of the security parameter $K$
$$Adv_A^3(K)=\left| Pr[Exp_A^3(K)=1]-\frac{1}{2} \right|$$

$B$ gets $(g, X, \hat{c}_1)$, $\hat{c}_1$ is the first component of the target ciphertext. Using the $A$ attack algorithm 3 as a subprogram, The following steps are taken to calculate $\hat{y}$, $\hat{c}_1=\hat{y} g \hat{y}^{-1}$. \\

$1.$ One chooses a random string $ \hat{h}\leftarrow_R \{0,1\}^{l(k)} $ as a guessing value for $H(\hat{Z})$, $B$ don't know the $\hat{Z}$, sents $pk=(g,X)$ to $A$; \\

$2.$ $H$ queries (Bounded polynomial times): $B$ build the list $H^{list}$ , element type is $( y, c_1, h )$,
initial value is $ (\ast, \hat{c}_1, \hat{h}) $, $\ast$ means unknow in this section. $A$ can query $H^{list}$ any time, $B$ respond as follows:(assume $A$ query $y$, $B$ compute $c_1=ygy^{-1}$)

$\bullet$ If $(y,c_1,h)$ in $H^{list}$, respond with $h$.

$\bullet$ If $(\ast,c_1,h)$ in $H^{list}$, respond with $h$, Replacing $(\ast,c_1,h)$  with $(y,c_1,h)$ in $H^{list}$.

$\bullet$ Else, choose random string $ h \leftarrow_R \{ 0, 1 \}^{l(k)} $, $h$ response, save the $ (y,c_1,h) $ into $H^{list}$.\\

$3.$ Decryption inquiries. $A$ queries $(\overline{c}_1, \overline{c}_2)$ to $B$, $B$ responds as follows:

If there is one item $(\overline{y}, \overline{c}_1, \overline{h} $ or $ \ast, \overline{c}_1, \overline{h}$ in the table, $D_{\overline{h}}(\overline{c}_2)$ response, else, choose a random string $ h\leftarrow_R \{ 0,1 \}^{l(K)}  $, $D_{\overline{h}}(\overline{c}_2)$ response, save $ \ast, \overline{c}_1, \overline{h}$ into $H^{list}$.

$4.$ Challenge. $A$ outputs two messages $ M_0, M_1 $, $B$ chooses $\beta\leftarrow_R \{0,1\}$, sets $ \hat{c}_2=E_{\hat{h}}(M_{\beta})$,
sends $( \hat{c}_1, \hat{c}_2 )$ (as the target ciphertext) to $A$;\\

$5.$ $A$ execution step 2, but he can't query $( \hat{c}_1, \hat{c}_2 )$.

$6.$ Guess. $A$ outputs $\beta'$, $B$ check the $H^{list}$, if exist one item $\hat{y}, \hat{c}_1, \hat{h}$, then output $\hat{y}$.

Assume that the $event \ D$: In the simulation, $H(\hat{Z})$ appears in the list $H^{list}$.

\subsubsection{Assertion 1} $B$ is complete in above simulation process. \\

\textbf{proof}: It is easy to know:

$\bullet$ In the $H$ inquiry of $A$, each value is answered by random string.

 In the real attack of $A$, the value of the function is generated by the random oracle, so the function value obtained by the $A$ is uniformly distributed;

$\bullet$  According to the structure of $H^{list}$, $\overline{h}=H(\overline{y}), \overline{c}_1 = \overline{y}g\overline{y}^{-1} $, so the decryption response of $B$ is valid.

Thus, both of the view of $A$ and its view in real attacks are not distinguishable in calculation.

\subsubsection{Assertion 2} In the simulation attacks above, $Pr[D]\geq 2\varepsilon$ \\

\noindent \textbf{proof} If $H(\hat{Z})$ does not appear in $H^{list}$, then $A$ have no $\hat{h}$, because of $\hat{c}_2=E_{\hat{h}}(M_{\beta})$ and $E$'s security, then $Pr[\beta'=\beta|\neg D] = \frac{1}{2}$. The rest is the same as Theorem 3. \\

In summary, if $A$ can take advantage of a non negligible advantage  $\varepsilon(K)$  win the IND-CCA2 game,
then $H(\hat{Z})$ appears at least in the probability of $2\varepsilon(K)$ in the $H^{list}$ in above simulation process, $B$ check the elements in $H^{list}$ one by one in step6. So the probability of the success of the adversary $B$ is equal to the event $D$, thus, $B$ attacks the CCS assumption with the non negligible advantage of $2\varepsilon(K) \ $. The algorithm 3 is secure against $IND-CCA2$.

\subsubsection{Note that} In this algorithm, we assume that the symmetric algorithm is IND-CCA2, because its construction method is already very mature.\cite{jour15}\cite{jour31}\cite{jour39}\cite{jour40}

\section{Conclusion}

For the first time, this paper uses a random oracle model to prove that the Ko cryptosystem\cite{jour1} is IND-CPA security and gives a non-hash public key cryptosystem on braid group, which is very similar to the ElGamal system\cite{jour30}. Finally, we propose an algorithm on braid group which is secure against chosen of ciphertext attack. This is a mixed encryption algorithm\cite{jour15}. The keys of the symmetric encryption part is produced by a random oracle. This design gets rid of the bondage of the braid group, and making the algorithm more compatible and more practical.

This paper opens the door of the research on the security of the braid cryptosystem, fills the blank of the research direction of provably security in braid cryptosystem, to effectively promote the algorithm to engineering applications, to a certain extent, this article has a pioneering spirit.

\subsubsection*{Acknowledgments.} We thank any reviewers to comments our paper.

\end{document}